\documentclass[10pt,twocolumn,letterpaper]{article}

\usepackage{ijcb}
\usepackage{times}
\usepackage{enumitem}

\usepackage[pdftex]{graphicx}
\graphicspath{{./images/}}
\DeclareGraphicsExtensions{.pdf,.jpeg,.png}

\usepackage{amsmath}
\usepackage{amssymb}
\usepackage{mathtools}
\usepackage[caption=false,font=footnotesize]{subfig}

\usepackage[pagebackref=true,breaklinks=true,letterpaper=true,colorlinks,bookmarks=false]{hyperref}

\ijcbfinalcopy 

\ifijcbfinal\fi

\newcommand{\B}[1]{\mathbf{#1}} 

\usepackage{fancyhdr}
\fancypagestyle{plain}{%
\fancyhead{} 
\fancyfoot{} 
\fancyfoot[C]{\textcopyright \space Callsign Inc. 2020. All Rights Reserved.\\\bigskip\thepage}
}

\begin{document}
%

\title{Swipe Dynamics as a Means of Authentication: \\Results From a Bayesian Unsupervised Approach}

\author{%
Parker Lamb\\Callsign, Inc.\and
Alexander Millar\\Callsign, Inc.\and
Ramon Fuentes\\Callsign, Inc.
}

\maketitle

\begin{abstract}

The field of behavioural biometrics stands as an appealing alternative to more traditional biometric systems due to the ease of use from a user perspective and potential robustness to presentation attacks. This paper focuses its attention to a specific type of behavioural biometric utilising swipe dynamics, also referred to as touch gestures. In touch gesture authentication, a user swipes across the touchscreen of a mobile device to perform an authentication attempt. A key characteristic of touch gesture authentication and new behavioural biometrics in general is the lack of available data to train and validate models. From a machine learning perspective, this presents the classic curse of dimensionality problem and the methodology presented here focuses on Bayesian unsupervised models as they are well suited to such conditions. This paper presents results from a set of experiments consisting of 38 sessions with labelled `victim' as well as blind and over-the-shoulder presentation attacks. Three models are compared using this dataset; two single-mode models: a shrunk covariance estimate and a Bayesian Gaussian distribution, as well as a Bayesian non-parametric infinite mixture of Gaussians, modelled as a Dirichlet Process. Equal error rates (EER) for the three models are compared and attention is paid to how these vary across the two single-mode models at differing numbers of enrolment samples.

\end{abstract}

\section{Introduction}

The advent of mobile computing and widespread availability of mobile devices calls for authentication systems that can be easily deployed to and used from any mobile device. While traditional biometric modalities such as facial and fingerprint identification offer high accuracy, these approaches present other difficulties, from preserving user privacy to the lack of availability of high quality sensors in common mobile computing devices. Furthermore, while certain biometric modalities such as facial and voice recognition can boast high accuracy when evaluated against random attackers, their performance can drop significantly when placed against targeted attacks such as simple presentation attacks \cite{mohammadi2017deeply}.

Behavioural biometrics offer an elegant alternative to traditional biometrics, the premise being that certain behaviours such as typing or web browsing possess enough characteristic information about an individual to identify them. Research in the field has now been active for about four decades, with modalities such as keystroke dynamics being investigated as far back as the 1970s \cite{forsen1977personal}. While the viability of the idea was established, early behavioural biometrics systems often required bespoke infrastructure and data acquisition, preventing general use. 

The behavioural biometrics field is currently undergoing a second wave of innovation as researchers and practitioners realise the potential the current interconnected world of cyber-physical systems brings. This has brought about the idea of mobile phone touch-based behavioural biometrics. The concept has been introduced in various different flavours, but the underlying idea is to utilise the available sensor data in modern mobile phones to characterise a user's behaviour for the purposes of authentication. Sensor channels may include touch screen coordinates, pressure, thumb size and alignment, as well as data from any embedded sensors, if these are available. Many mobile devices are equipped with sensors that capture acceleration, rate of rotation, orientation, and magnetometer readings.

\subsection{Continuous and non-continuous authentication}
The field of touch-based mobile phone biometrics can be broadly categorised into continuous and non-continuous authentication. Continuous authentication implies that mobile phone data is gathered and monitored throughout a mobile browsing session. Arguably, the continuous authentication problem has received the most attention so far \cite{mondal2015swipe,kumar2016continuous,putri2016continuous}.

In the case of non-continuous authentication, behavioural touch data is only gathered as a user  performs the authentication attempt. While this approach has been investigated in \cite{tse2019behavioural,antal2016swipe}, both of these studies concern touchscreen data that is gathered from users in a passive manner and subsequently used in order to evaluate the accuracies of different modelling strategies. In \cite{antal2016swipe}, touchscreen data from users is collected as they fill out a questionnaire on an Android mobile device. In their experiments, they show that authentication accuracy can be improved dramatically if multiple swipe attempts are included in the models and quote equal error rates (EER) of less than 0.5\% in such settings.

A noteworthy point highlighted in these previous studies is that the accuracy of biometric modalities which incorporate swipe dynamics tends to increase as more data is gathered. Gathering  data is easier in a continuous authentication mode compared to the non-continuous case. This means that it is relatively easier to build a `template' of a user's swipe dynamics given a single session. However, this might result in  the user's template having much higher variance, thus making it more prone to attacks. The results from studies published so far must be interpreted with this in mind. As far as the authors are aware, no study exists that explicitly incorporates data from over-the-shoulder (OTS) impostors, where the impostors are assumed to have observed the genuine user's touchscreen behaviour. The lack of such studies is partly what motivates the current paper.

\subsection{Swipe dynamics as a means of authentication}
The appeal of behavioural biometrics as authenticators is their ease of use, particularly in the case of touch screen swipe dynamics, which translates to low user friction while remaining strong against different types of attacks.

In this paper, the key problem under consideration is that of utilising swipe dynamics as a means of non-continuous authentication. The subsequent investigation identifies techniques that are able to instantly assess whether a touchscreen swipe pattern belongs to the individual in question, minimising risk of an impostor accessing the system at any point. A simple authenticator that requires few samples of behaviour before a user can be enrolled into the system is generally desirable. An example of how such an authenticator would look in practice from a user's perspective is shown in Section \ref{sec_data_acquisition}. The non-continuous authentication problem has not been systematically studied to the authors' knowledge.

\subsection{Contributions of this paper}
Bayesian novelty detection is introduced as a means of enabling swipe dynamics authentication. The Bayesian approach alleviates several inherent problems of the data, particularly high dimensionality and low sample sizes. The ability to set prior beliefs on model parameters allows accurate predictions to be made for what would otherwise be an ill-posed problem. Additionally, the use of Bayesian nonparametrics makes it possible to infer multiple distributions from the data when they exist, in a manner that naturally avoids overfitting the probability distribution of the data.

Results are presented on a new experimental dataset collected in a non-continuous context which contains negative samples from both blind and over-the-shoulder impostors for a simple swipe task. To the authors' knowledge model performance has not been evaluated against active over-the-shoulder impostors elsewhere in the field for such a task.

\section{Modelling Swipe Dynamics}

Machine learning models have become a strong focus of recent research efforts in biometric authentication, and behavioural biometrics is no exception. One important point that distinguishes behavioural modalities from others is the lack of availability of public data, which significantly affects the appropriate choice of statistical and machine learning models that are most suitable for the task. Modalities such as facial and voice verification enjoy the availability of many public datasets which can be used to develop and evaluate modelling strategies \cite{FIERREZ20071389,10.1007/BFb0016021,10.1007/3-540-44887-X_98}.

While several public datasets concerning continuous swipe authentication exist \cite{Mahbub_2016, Frank_2013, ANTAL20157}, this data cannot readily be used for developing non-continuous models as it is passively collected while users interact with the device, rather than through a controlled authentication task. Other studies have published datasets that include touchscreen data from non-continuous authentication for complex tasks, such as drag-and-drop \cite{alej2020becaptcha} or drawing digits \cite{Tolosana2019}. Neither of these datasets contain samples from active impostors.

The general lack of available data for simple swipe biometrics, combined with the fast evolution of the data acquisition quality of mobile computing devices makes \textit{novelty detection} \cite{pimentel2014review}, an unsupervised method, a suitable modelling strategy for mobile phone touch-based authentication. Supervised models are difficult to utilise in a real-world environment as negatively labelled samples are not available in sufficient quantities per-user. For this reason, most of the previous work has focused on the use of novelty detection as a general strategy. However, a common factor amongst recent research is a `scatter-gun' approach to simply reporting EERs for well-known novelty detection methods \cite{pimentel2014review}, including One-Class Support Vector Machines, neural networks, K-Nearest-Neighbours and Gaussian mixture models \cite{guest2020swipe}.

In this study, focus is given to a class of novelty detection methods that model the probability distribution of the behavioural features. A nuanced discussion is contributed to how these models solve certain practical challenges; namely that of training with few samples without overfitting as well as the ability to capture multiple behaviours exhibited by a single user in the case of the mixture model. These arguments are presented in Section \ref{sec_bayesian_novelty_detection}.

\subsection{Features from swipe gestures}

Arguably, the feature space derived from swipe data is as important as the modelling strategy. A useful feature space is one that is consistent for a given user and discriminating of impostor behaviour. It is not known a-priori which features are consistent for any given user. The strategy adopted by previous studies has been to use a range of summarising features from each swipe gesture. It is common to compute features such as horizontal and vertical coordinates, velocity, angles, first and second order moments, and start and end points. 

The feature vectors used in this study were derived from the raw touch screen coordinates and pointer size measurement. The feature vectors were assembled using the horizontal and vertical coordinates together with pointwise velocity, acceleration, angle, and angular acceleration. These coordinates were resampled, using techniques borrowed from functional data analysis, to a fixed-length vector representing the swipe position, velocities and accelerations on an even time grid representing one swipe `cycle'. The touch pointer size was also resampled accordingly and used as an additional feature. An illustrative example of raw swipe data is shown in the following section.

\subsection{Enrollment and user behaviour drift}
When users have not authenticated enough times for a model to be trained on their behaviour they are considered to be in the `enrolment' phase. A behavioural authenticator should be able to begin making predictions with as few samples as possible, ideally in the range of one to five. This low threshold facilitates ease-of-use and protects user accounts from unauthorised access as quickly as possible. After the enrolment threshold has been passed the user will have a valid trained model and predictions can be made. Outside of laboratory conditions a behavioural model can be retrained at regular intervals using newer data as users continue to authenticate. Frequent retraining combats `behavioural drift' as users become more familiar with the system. While these two notions of continuous learning and behavioural drift each merit a nuanced discussion, this is left outside of the scope of the current paper. 

\begin{figure}[!t]
  \centering
  \includegraphics[width=0.6\linewidth]{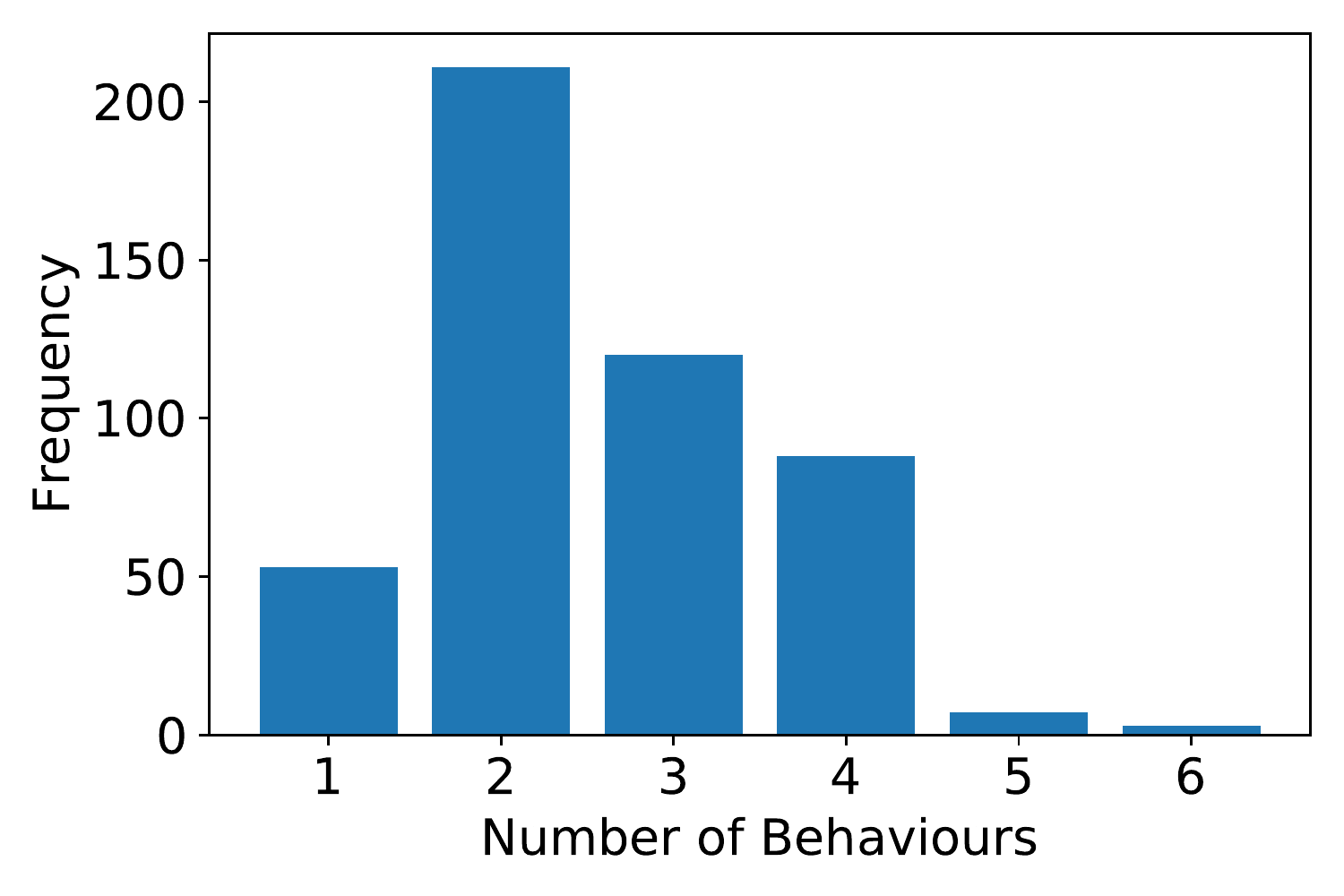}
  \caption{Approximate Distribution of User Behaviours}
  \label{fig_user_behav_dist}
\end{figure}

Over time, a significant portion of users tend to develop multiple well-defined behaviours. These behaviours are usually maintained simultaneously and can have explicit meanings such as a user swiping with two different hands, or simply be subtle unconscious changes in swiping patterns. This motivates the need to model them explicitly. As an illustrative example, Figure \ref{fig_user_behav_dist} shows the distribution of the number of behaviours observed from a sample of touch gesture behavioural biometrics users, in which it is evident that the majority exhibit more than one behaviour.

\begin{figure}[!t]
    \begin{center}
        \subfloat[Single Behaviour]{\includegraphics[width=0.5\linewidth]{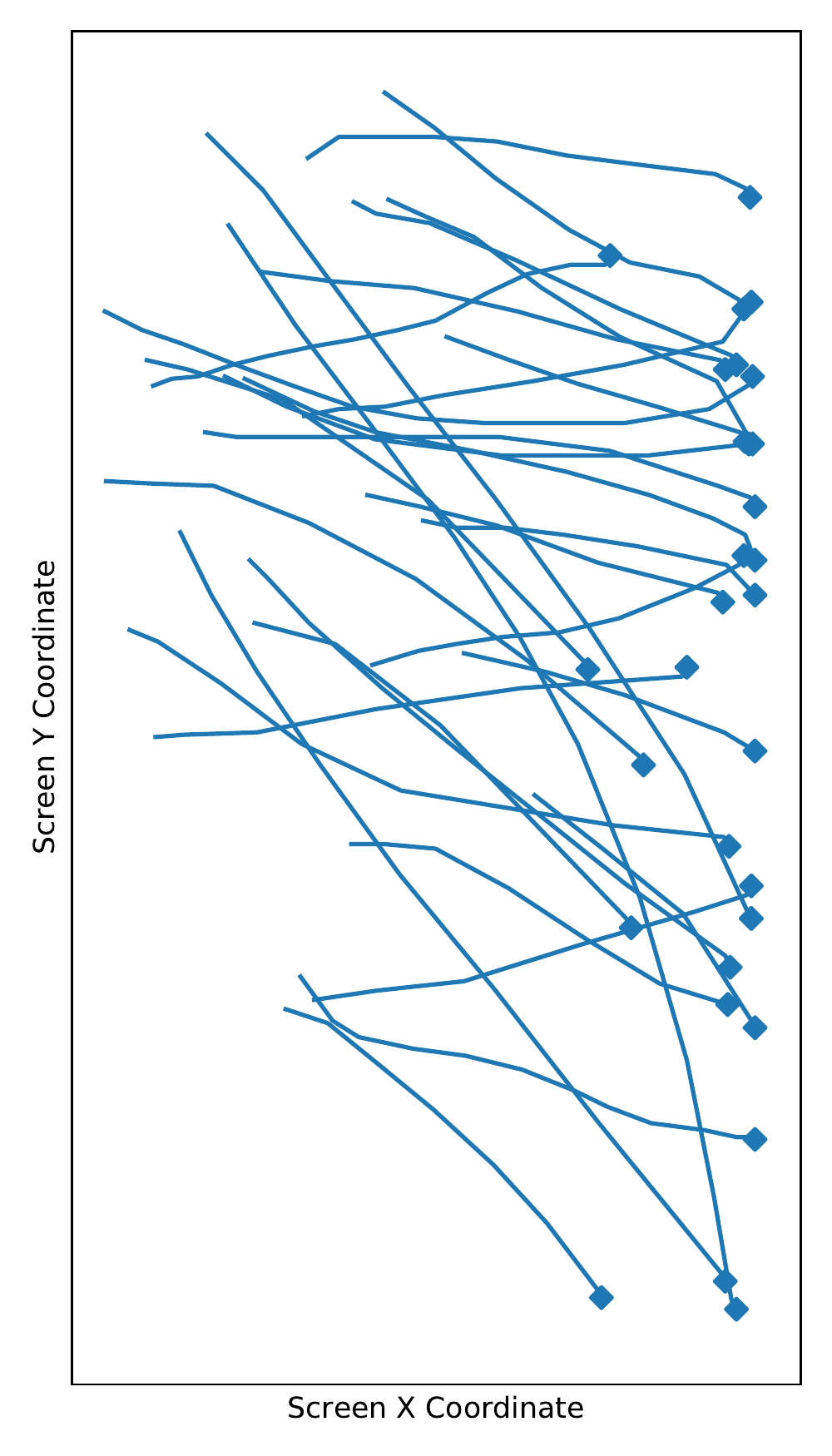}%
        \label{fig_swipe_single_behav}}
        \hfil
        \subfloat[Multiple Behaviours]{\includegraphics[width=0.5\linewidth]{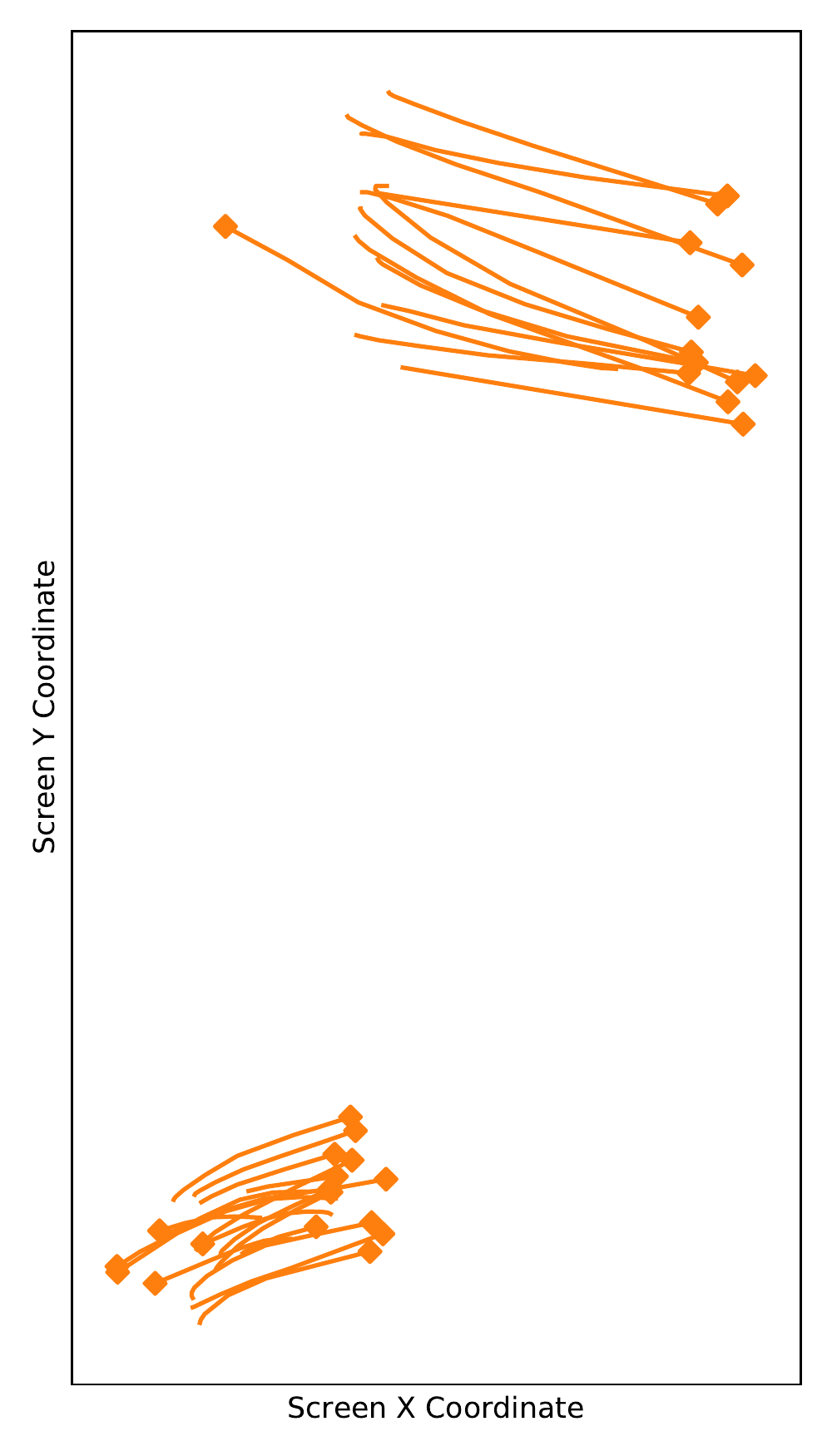}%
        \label{fig_swipe_multi_behav}}
    \end{center}
    \caption{Touch Screen Gesture Behaviours}
    \label{fig_swipes_example}
\end{figure}

Figure \ref{fig_swipes_example} shows examples of users who exhibit single and multiple behaviours. The lines in these plots denote the paths of swipe gestures, with the marker indicating the end point of the gesture. A single behaviour is defined by a unified grouping of gestures which have similar direction and shape, as can be seen in Figure \ref{fig_swipe_single_behav}. Multiple behaviours occur when two or more single behaviour groups are present for a single user. These groups can overlap, as seen in Figure \ref{fig_swipe_multi_behav}, where the lower-left cluster consists of two behaviour groups with opposing directions.

\subsection{Bayesian novelty detection}
\label{sec_bayesian_novelty_detection}

Novelty detection is an unsupervised method that deals with the problem of determining whether a new observation belongs to the class of previously observed data. Given the small number of observations typically found in non-continuous swipe dynamics authentication, coupled with the inherent high dimensionality of time-series data, the problem is classically ill-posed. With few training samples the ability to train models directly from the available behaviour of a single user is greatly diminished. The use of classical machine learning models in this instance will typically result in inaccurate predictions due to the ill-posed nature of the problem and lack of variance in the single-class training data.

This motivates the use of Bayesian models where information from priors can be incorporated into the model. This has the effect of increasing the total amount of information available to the model and allows more accurate predictions to be made. Such priors can be derived from a reasonable default or the general population of users, if available. Taking a Bayesian approach also yields a probabilistic estimate of model parameters, therefore alleviating overfitting.

The central approach taken to modelling presented here is based around fitting probability distributions to the given training data. Scoring of subsequent samples is done via the model likelihood, where the decision boundary is calibrated based on the likelihood of the training samples.

Three models are presented, representing an evolution of complexity in modelling approach. The first model is a non-Bayesian multivariate Gaussian distribution with a regularised covariance matrix. The second and third models are both Bayesian; a multivariate Gaussian distribution with priors derived from the user population and an infinite Gaussian mixture with a Dirichlet process prior on the components \cite{rasmussen2000infinite}. The relative performance of these models in the swipe dynamics task is presented in Section \ref{sec_results}.

\subsubsection{Shrunk Covariance}

The shrunk covariance model is the simplest of the three presented in this paper. It models the probability distribution of the data, $p(\B{x} ; \theta)$ as a multivariate Gaussian, $\B{x} \sim \mathcal{N}(\B{\mu}, \B{\Sigma})$, which is parametrised by its mean and covariance $\theta = (\B{\mu},\B{\Sigma})$. 

When the number of samples is large relative to the number of features, the maximum likelihood estimate (MLE) provides a good representation of the underlying data covariance structure. However, when the number of samples is comparatively small against the number of dimensions, the covariance MLE is a poor estimate of the true underlying data covariance. Maximum likelihood estimation of $\B{\Sigma}$ in this setting is well known to be unstable and leads to singularities when inverting it to arrive at the precision. One way in which this instability manifests itself is by adding extra variance.   

A common solution to alleviate this is to \textit{shrink} the estimated covariance matrix towards the diagonal,
\begin{equation}\label{eq:shrunk_cov}
    \B{\Sigma}_{\textrm{shrunk}}(\alpha) = \alpha \B{\hat{\Sigma}} + (1-\alpha) \textrm{diag}(\B{\hat{\Sigma}})
\end{equation}
where $\alpha$ is a hyperparameter that controls the strength of the covariance shrinkage. The shrunk estimate of Equation \eqref{eq:shrunk_cov} thereby reduces the overall variance of the covariance estimate. This is a form of regularisation in covariance esitmation. Selecting an appropriate value for $\alpha$ can be be done via cross-validation.

Prediction in the model is carried out by evaluating the model log-likelihood of the Gaussian distribution with parameters  $(\B{\mu}, \B{\Sigma}_{\textrm{shrunk}})$, where $\B{\mu}$ is the maximum likelihood estimate of the mean, of a test sample  $\B{x}^*$. Alternatively, one could use the squared Mahalanobis distance of the test sample under the shrunk covariance estimate. 

\subsubsection{Bayesian Multivariate Gaussian}
\label{sec_bayesian_multivariate_gauss}

The shrunk covariance estimator can be considered a special case of estimation of the full posterior probability distribution of $\B{\mu}$ and $\B{\Sigma}$, given data $\mathcal{D}$. This can be carried out using Bayesian inference, which leads to the Bayesian multivariate Gaussian model.  Instead of using maximum likelihood estimates for the mean and covariance, Bayesian priors are constructed for these parameters. The priors can be defined by specifying a suitable belief on the parameter values (for example, position variance to be 15\% of screen size) in absence of any reliable observations, or alternatively it can be based on a sample of the user population.

For a single user, the sample distribution is assumed to be multivariate normal with conjugate (normal-inverse-Wishart) priors on the mean and covariance:
\begin{align}
p(\B{\Sigma}) &= \mathrm{InvWish}(\Psi_{0},\nu_{0})\\
p(\B{\mu} | \B{\Sigma}) &= \mathcal{N}(\mu_{0},k_{0}^{-1} \B{\Sigma})\\
p(\B{X}| \B{\mu}, \B{\Sigma}) &= \mathcal{N}(\B{\mu}, \B{\Sigma})
\end{align}

As the normal-inverse-Wishart prior on $(\B{\mu}, \B{\Sigma})$ is conjugate for a multivariate Gaussian likelihood, the posterior is of the same family, that is,
\begin{align}
    p(\B{\Sigma} | \mathcal{D}) & = \mathrm{InvWish}(\Psi_{N},\nu_{N})\\
    p(\mu | \B{\Sigma}, \mathcal{D}) & = \mathcal{N}(\mu_{N},k_{N}^{-1} \B{\Sigma})
\end{align}
where $\mathcal{D}$ denotes training data. The formulas determining the posterior parameters are closed-form and can be found in \cite{murphymlprob12}.

Prediction of a test sample $\B{x}^*$ is carried out using the model posterior predictive distribution, which has a closed-form analytic solution,
\begin{equation}
    p(x^*|\mathcal{D}) = t_{v_{N}-d+1}(\mu_{N}, \frac{k_N+1}{k_N(\nu_N -d+1)} \Psi_{N}) 
\end{equation}

Inferring the mean is relatively straightforward; one option is to let $k_{0} \to 0$, which loosely corresponds to an uninformative prior for $\mu$. Inference about the covariance is more challenging, given the comparatively high number of features in relation to training samples, which can be on the order of 5-10 times. A simple approach is to utilise the unbiased estimate of the pooled (over the profiles) covariance $\B{\hat{\Sigma}}_{\textrm{pooled}}$ in the construction of the prior,
\begin{equation}
    \Psi_0 = (\nu_{0} - d - 1)\B{\hat{\Sigma}}_{\textrm{pooled}}
\end{equation}

This prior has the effect of shrinking the posterior distribution towards the pooled covariance estimate, where the degrees of freedom controls the level of shrinkage.

In the results presented in this paper, a shrunk estimate of the pooled covariance was used:
\begin{equation}
    \B{\hat{\Sigma}}_{\textrm{shrunk pooled}}(\alpha) = \alpha {\B{\hat\Sigma}}_{\textrm{pooled}} + (1 - \alpha) \textrm{diag}(\B{\hat{\Sigma}}_{\textrm{pooled}})
\end{equation}
where $\alpha$ determined by cross validation, using the log-likelihood of the normal distribution evaluated at ($\bar{x}_p, \B{\hat{\Sigma}}_{\textrm{shrunk pooled}}(\alpha)$) for profile $p$ as a scoring criterion.

Intuitively, the advantage of this approach is that when the number of samples is small, this method allows us to borrow strength from other profiles in inferring the covariance, and converges on the profile's true covariance as the number of samples increases.

\subsubsection{Infinite Gaussian Mixture}

The infinite Gaussian mixture is a Bayesian nonparametric model, meaning that the parameter space has infinite dimensions, which allows the parameter space to grow as the parameter space is explored and more data is observed. In practice this means that the model is very expressive, as it allows the size of the parameter space to match the amount of observed data.  The idea behind this model in the context of swipe dynamics is to explicitly model for multiple distributions as opposed to the previous models which assume a single Gaussian. This is intended to cater for situations where the user has developed multiple behaviours. It is also a more flexible approach than using a traditional Gaussian mixture, as the number of components can be inferred from the data in a principled manner. The topology of the input feature space is assumed to be such that the behaviour groups are isolated.

The number of mixture components is chosen via the Chinese Restaurant Process \cite{gelmanbda04}, which iteratively assigns training samples to components based on each sample's likelihood of belonging to a particular component. Additional components are generated when the likelihood of a sample belonging to a hypothetical new component is higher than that of any existing ones.

The key equations for this model are Equations \eqref{eqn_multi_cluster_prob} and \eqref{eqn_multi_new_cluster_prob}. These equations determine the likelihood of a sample belonging to a particular component or being assigned to a new component, respectively. All training samples are assigned to a single component initially and are reassigned to the component with the greatest likelihood as the algorithm progresses. After a suitable number of sampling iterations the component assignments for each sample will have converged. After convergence, a decision boundary for this model can be constructed equivalently to the models above.
\begin{equation}
  \label{eqn_multi_cluster_prob}
  \begin{multlined}
    p(c_i|\textbf{c}_{-i}, \mu_k, \tau_k, \alpha) \propto\\
    \frac{n_{-i, k}}{n - 1 + \alpha} \mathcal{N}\left(\tilde{x}_i; \frac{\bar{x}_k n_k \tau_k + \mu_0 \tau_0}{n_k \tau_k + \tau_0}, \frac{1}{n_k \tau_k + \tau_0} + \sigma^2_y \right)\rule[-1.5em]{0pt}{0pt}
  \end{multlined}
\end{equation}
\begin{equation}
  \label{eqn_multi_new_cluster_prob}
  \begin{multlined}
    p(c_i \neq c_k, \forall j \neq i | \mathbf{c}_{-i}, \mu_0, \tau_0, \alpha) \propto\\
    \frac{\alpha}{n - 1 + \alpha} \mathcal{N}(\tilde{x}_i; \mu_0, \sigma^2_0 + \sigma^2_x)
  \end{multlined}
\end{equation}

The hyperparameters for the model are \(\alpha\), \(\mu_0\), \(\sigma^2_0\), and \(\sigma^2_y\). These are the \(\alpha\) parameter of the Dirichlet distribution, prior component mean, prior component variance, and a prior on noise in the data, respectively. Note that \(\mu_0 \in \mathbb{R}^N\) and \(\sigma^2_0, \sigma^2_y \in \mathbb{R}^{N \times N}\). As can be seen from the equations above, the parameter \(\alpha\) controls the model's overall propensity to generate new clusters. The choice of the noise prior, \(\sigma^2_y\), also influences cluster generation, with more assumed noise leading to fewer clusters. The prior distributions on the component means and covariances are constructed identically to those in Section \ref{sec_bayesian_multivariate_gauss}, as each component is a multivariate Gaussian. The components all share identical prior distributions.

The remaining parameters in Equations \eqref{eqn_multi_cluster_prob} and \eqref{eqn_multi_new_cluster_prob} are component assignments \(c_i\), the number of samples assigned to a component \(n_i\), component precisions \(\tau_k\), and the total number of training samples \(n\). A negative subscript indicates all components excluding the one indicated. The overall structure of this model was derived from \cite{li2019tutorial}, additional detail can be found there.

\section{Experiments}
\label{sec_experiments}

One of the key motivations behind this study is the evaluation of swipe authenticators under realistic attack scenarios. The two attack scenarios of interest are \textit{blind attacks}, where the attacker is assumed to never have observed a victim's behaviour and \textit{over-the-shoulder} (OTS) attacks, implying that the attacker has been allowed to observe the victim's behaviour by looking over their shoulder. Evaluating model performance for a swipe authenticator under these scenarios required gathering of fresh experimental data, given the lack of suitable public benchmarks that consider this problem specifically. The experimental setup and data collection used for the results presented in this paper is described in this section. 

\subsection{Data acquisition and authentication prompt}
\label{sec_data_acquisition}

Swipe data was collected through a specially designed mobile phone application. The application collects touch screen and embedded sensor data, though only the touch data was used in this study.

\begin{figure}
  \centering
  \includegraphics[width=0.5\linewidth]{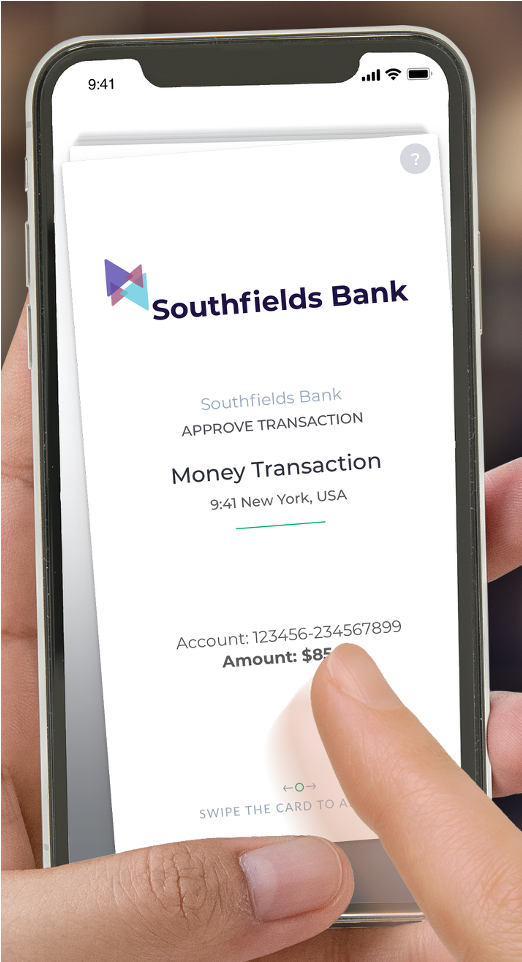}
  \caption{Authentication Prompt}
  \label{fig_swipe_card}
\end{figure}

The application prompts users to authenticate for a secure session by using a swipe gesture to slide a dialog off the screen. The user interaction with the mobile device is therefore relatively constrained, as users are encouraged by the interface to swipe the dialog horizontally and must use gestures of a minimum length to remove it. Additionally, the application is designed such that the screen orientation is forced to portrait mode. These constraints facilitate more accurate authentication as users must interact with the application in a controlled manner, even outside of laboratory conditions. A depiction\footnote{N.B. `Southfields Bank' is fictional and the corresponding branding was created for the demo application to give users context.} of the authentication prompt dialog can be found in Figure \ref{fig_swipe_card}. The user has begun swiping the dialog to the right in this figure.

The experiments were carried out using a range of mid- to high-end Android devices with a mix of device brands and models. Therefore the majority of the sessions were each carried out with a different device type which introduces some variance in the quality of the data. Most of the devices had a sample rate between 60-200Hz, with one outlying device sampling at approximately 500Hz. The application was designed to collect touch screen data at the highest possible sampling rate for each device. 

\subsection{Experimental procedure}

The experiments described in this paper were conducted over three separate days (Oct 23\textsuperscript{rd}, Oct 30\textsuperscript{th} and Nov 20\textsuperscript{th} 2019). On each day, a cohort of subjects was assembled and split into groups of three, where each member of the group assumed the roles of either victim, blind attacker, or OTS attacker. There was no special selection process for the subjects, they were gathered from volunteers in an office environment. No demographic data was collected for the subjects, however their age range varied from early twenties to early sixties. 

Each data collection session followed the following sequence:
\begin{enumerate}[itemsep=0.3em,topsep=0.3em]
    \item Victim performs $N$ authentication attempts.
    \item Blind attacker performs $N$ authentication attempts.
    \item Victim performs $N$ authentication attempts, where the OTS attacker gets to observe the victim's swipe behaviour. 
    \item OTS attacker performs $N$ authentication attempts. 
\end{enumerate}

\noindent where the value of $N = 10$ was used for the Oct 23 and Nov 20 sessions, and $N = 20$ for the Oct 30 session. In total, 38 sessions were recorded.

Each subject was given a basic explanation of the authentication process and given the opportunity to perform a small number of trial swipe attempts before any data was gathered. Subjects were free to sit or stand, most chose to stand. Data on subject position was not recorded. 


\section{Results}
\label{sec_results}

\subsection{Experiment Evaluation}
In all cases the models were trained on the first 10 genuine samples (sorted chronologically) for each user. A number of the profiles in the dataset only contain 20 samples total for the genuine user so the training sample size was restricted to keep the evaluation balanced. The models were evaluated on the remainder of the genuine samples not used for training and all of the samples for the given impostor.

The metric used for evaluation is the Equal Error Rate (EER). The EER is evaluated on a per-user basis as this provides insight into the way the errors distribute across the population of users in the study. Calculating the EER on a global basis can hide poor performance for users which are difficult to classify. The statistics presented in Table \ref{tab_results} are the means and medians of these user EER distributions.

Of the 38 profiles in the dataset, 6 failed to enroll due to lack of sufficient genuine samples. A number of the swipe gesture samples for these profiles were rejected due to insufficient data quality. The results presented later in this section therefore only concern the 32 remaining profiles.

\subsection{Experiment Results}

\begin{table}[!t]
    \renewcommand{\arraystretch}{1.2}
    \begin{center}
        \begin{tabular}{ccccc}
            \hline\hline
            \multicolumn{1}{c}{\textbf{Model}} & \multicolumn{2}{c}{\textbf{Blind EER}} & \multicolumn{2}{c}{\textbf{OTS EER}} \\
            & Mean & Median & Mean & Median \\
            \hline
            Shrunk Covariance & 5.07 & 0.00 & 16.06 & 9.55 \\
            Bayesian Gaussian & 4.54 & 0.00 & 16.10 & 5.96 \\
            Infinite Mixture & 4.99 & 0.00 & 15.70 & 9.27 \\
            \hline\hline
        \end{tabular}
    \end{center}
    \caption{Results}
    \label{tab_results}
\end{table}

\begin{figure}[!t]
    \captionsetup[subfigure]{justification=centering}
    \captionsetup[subfloat]{farskip=0pt}
    \begin{center}
        \subfloat[Shrunk Covariance, Blind Impostor][Shrunk Covariance,\\ Blind Impostor]{\includegraphics[width=0.5\linewidth]{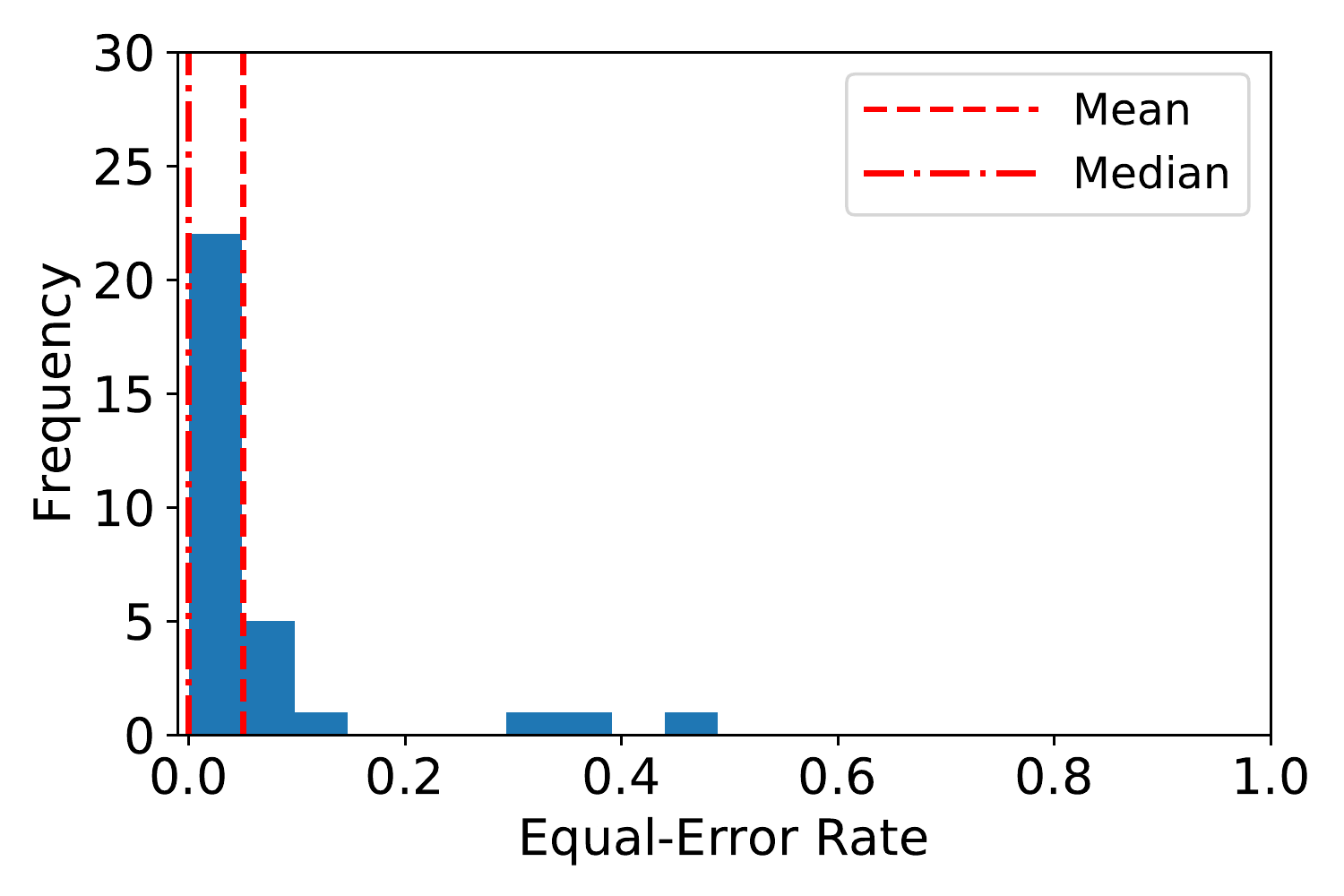}%
        \label{subfig_shrunk_cov_blind_imp_score_dist}}
        \hfil
        \subfloat[Shrunk Covariance, OTS Impostor][Shrunk Covariance,\\ OTS Impostor]{\includegraphics[width=0.5\linewidth]{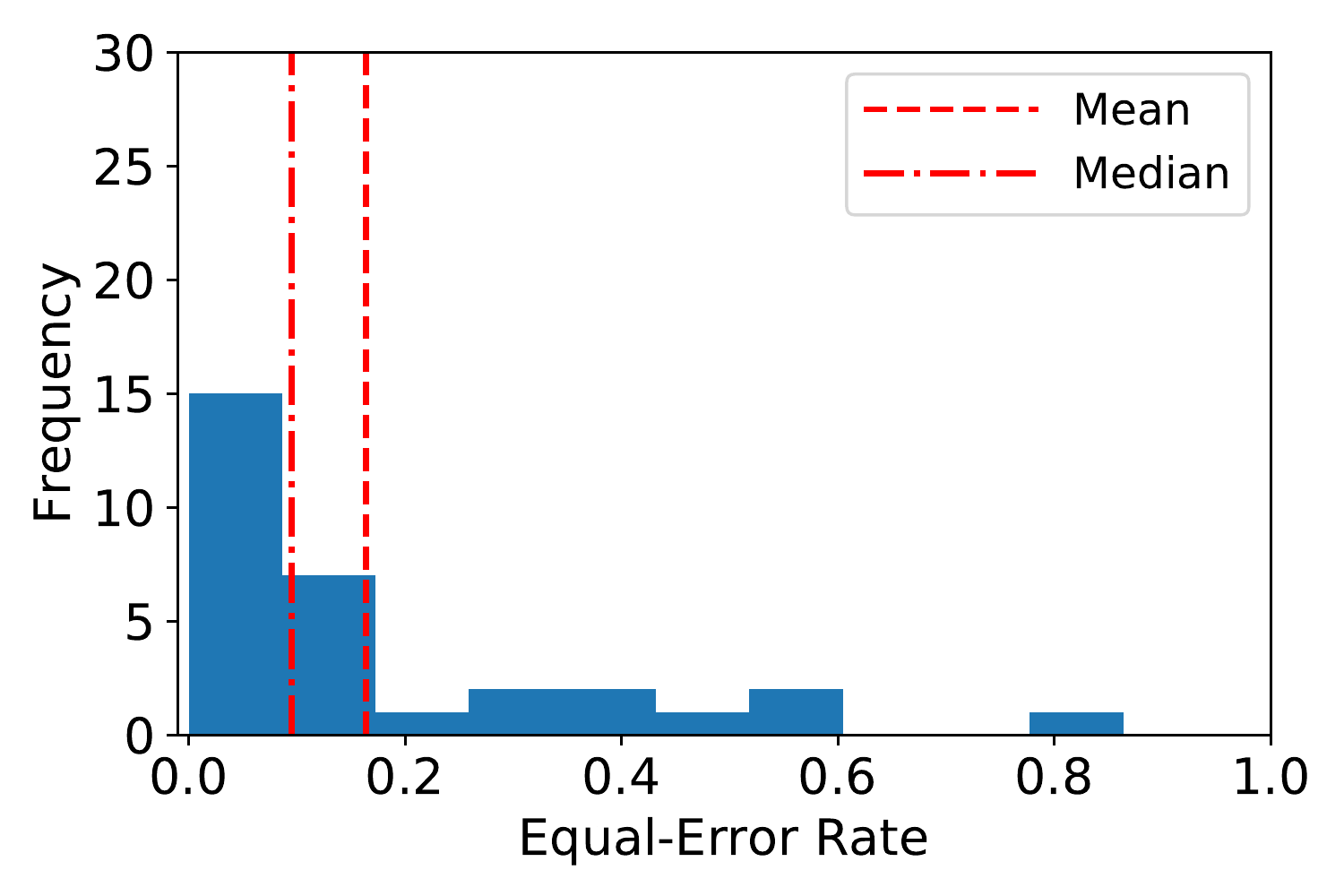}%
        \label{subfig_shrunk_cov_ots_imp_score_dist}}\\
        \subfloat[Bayesian Gaussian, Blind Impostor][Bayesian Gaussian,\\ Blind Impostor]{\includegraphics[width=0.5\linewidth]{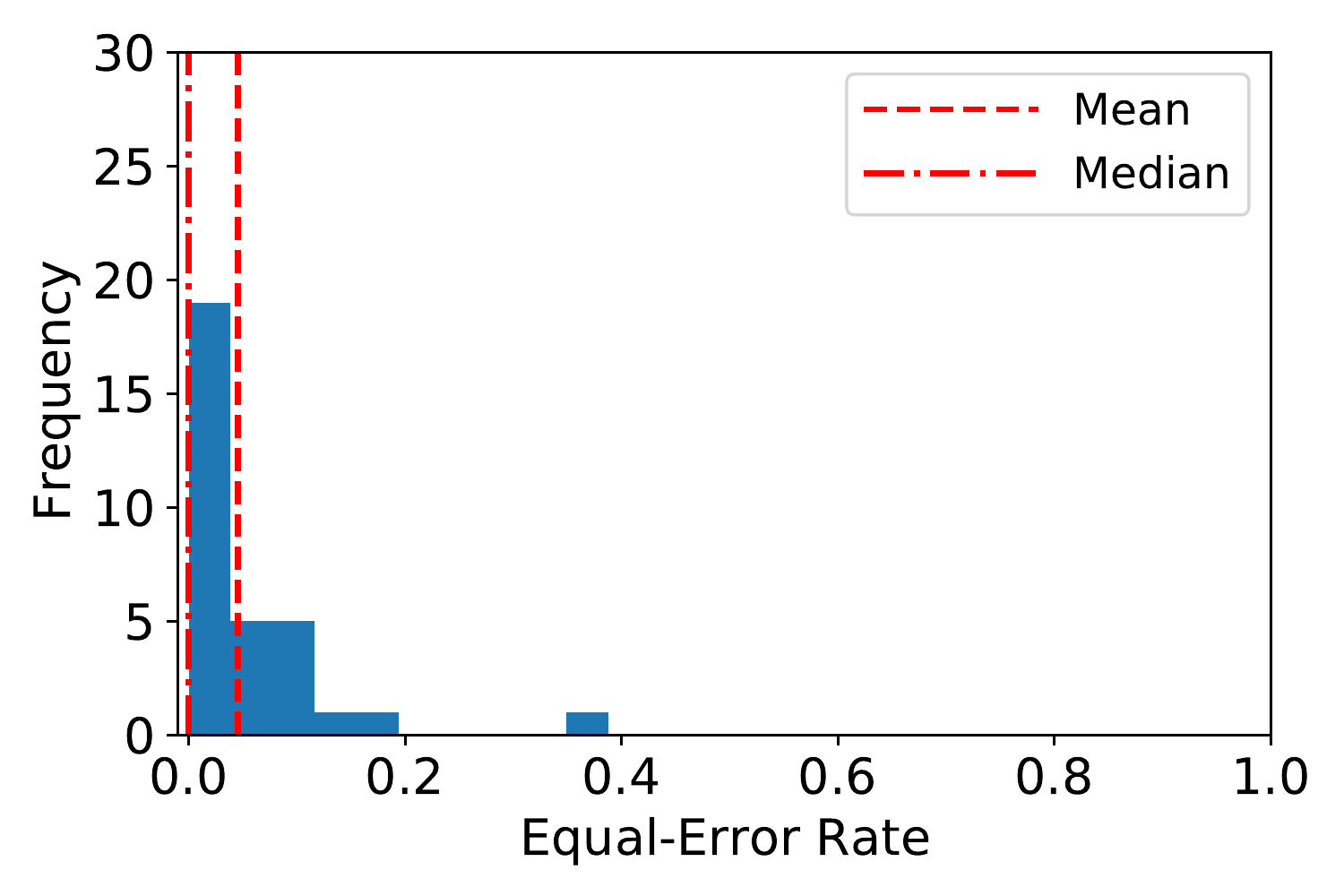}%
        \label{subfig_bayes_gauss_blind_imp_score_dist}}
        \hfil
        \subfloat[Bayesian Gaussian, OTS Impostor][Bayesian Gaussian,\\ OTS Impostor]{\includegraphics[width=0.5\linewidth]{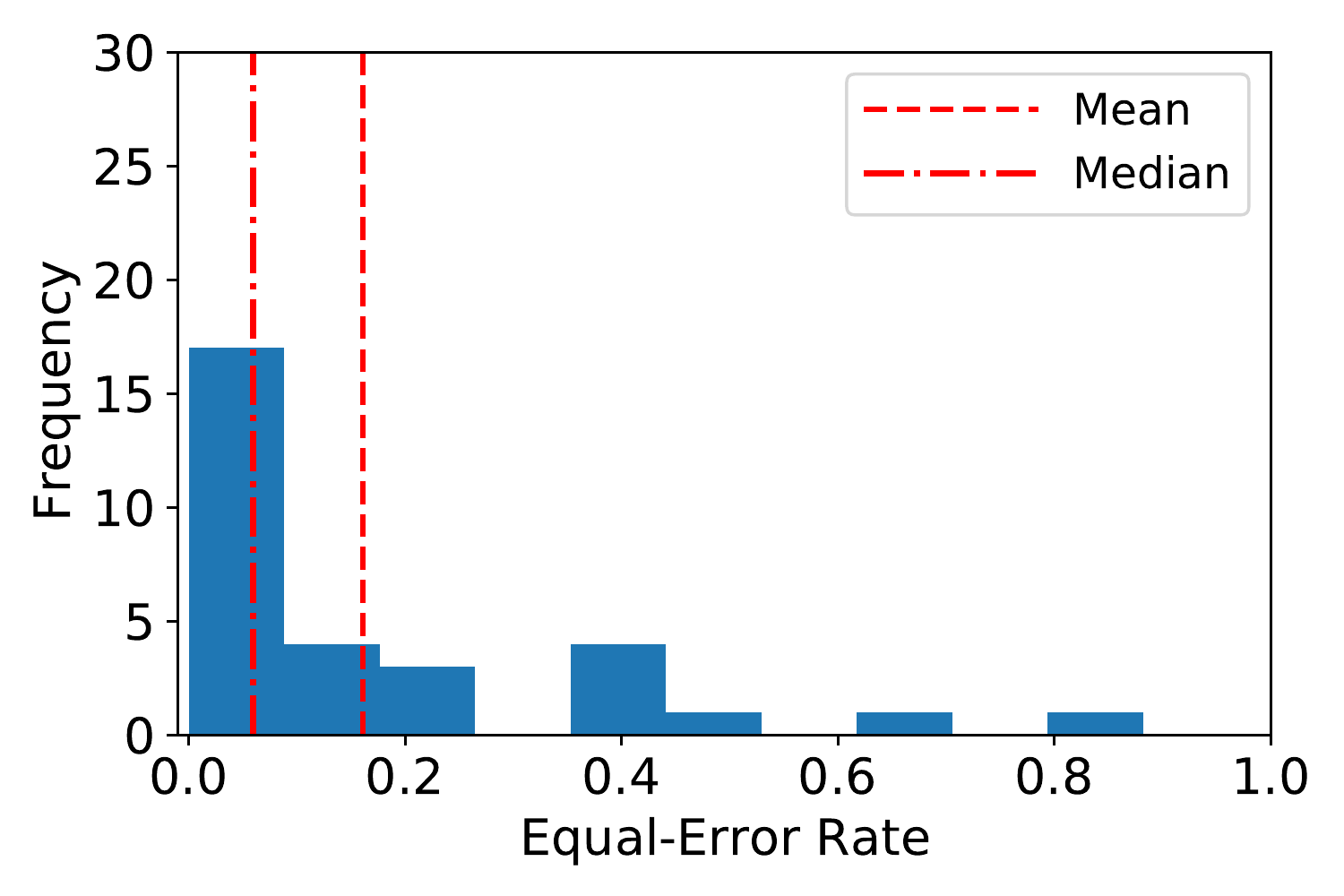}%
        \label{subfig_bayes_gauss_ots_imp_score_dist}}\\
        \subfloat[Infinite Mixture, Blind Impostor][Infinite Mixture,\\ Blind Impostor]{\includegraphics[width=0.5\linewidth]{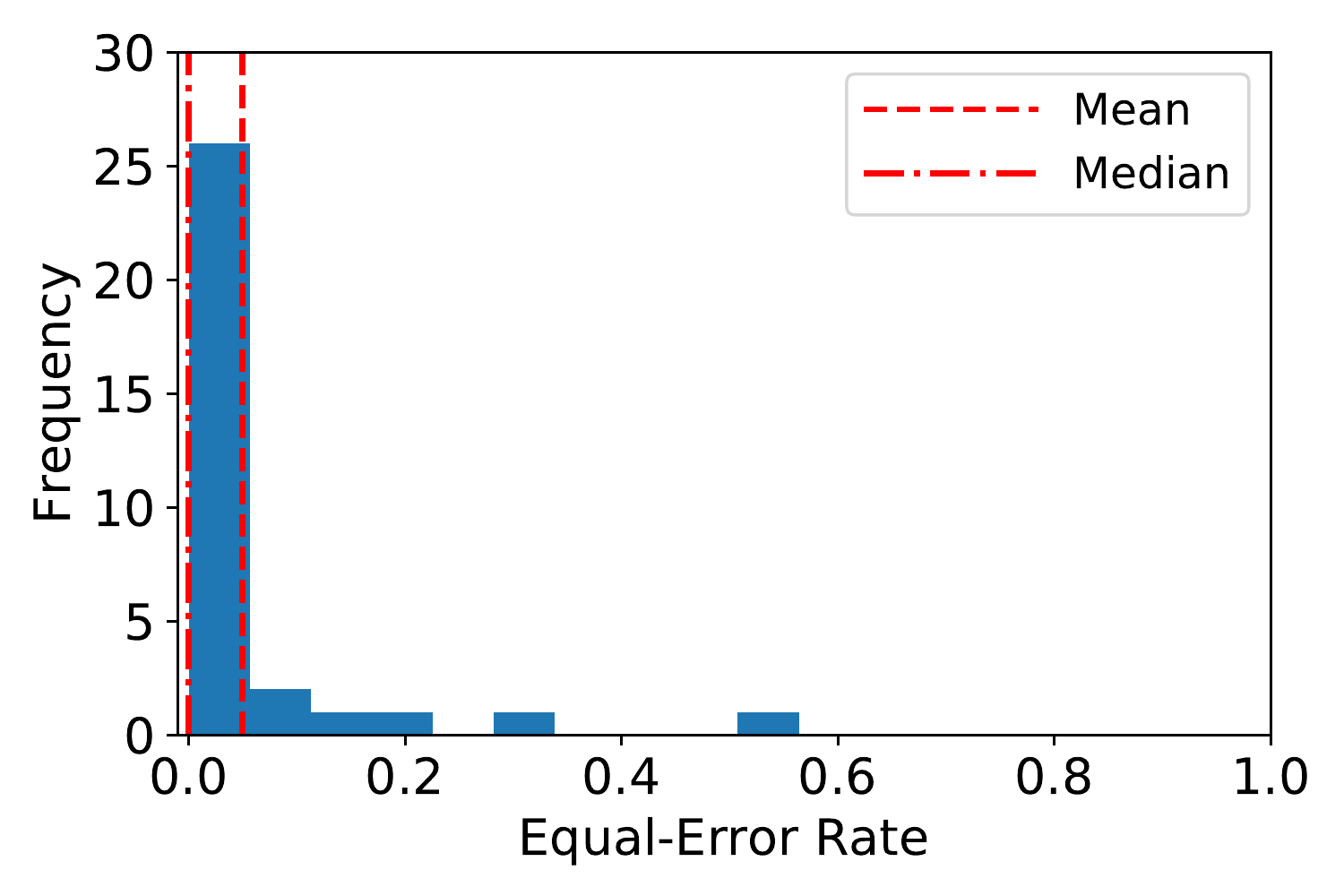}%
        \label{subfig_inf_mix_blind_imp_score_dist}}
        \hfil
        \subfloat[Infinite Mixture, OTS Impostor][Infinite Mixture,\\ OTS Impostor]{\includegraphics[width=0.5\linewidth]{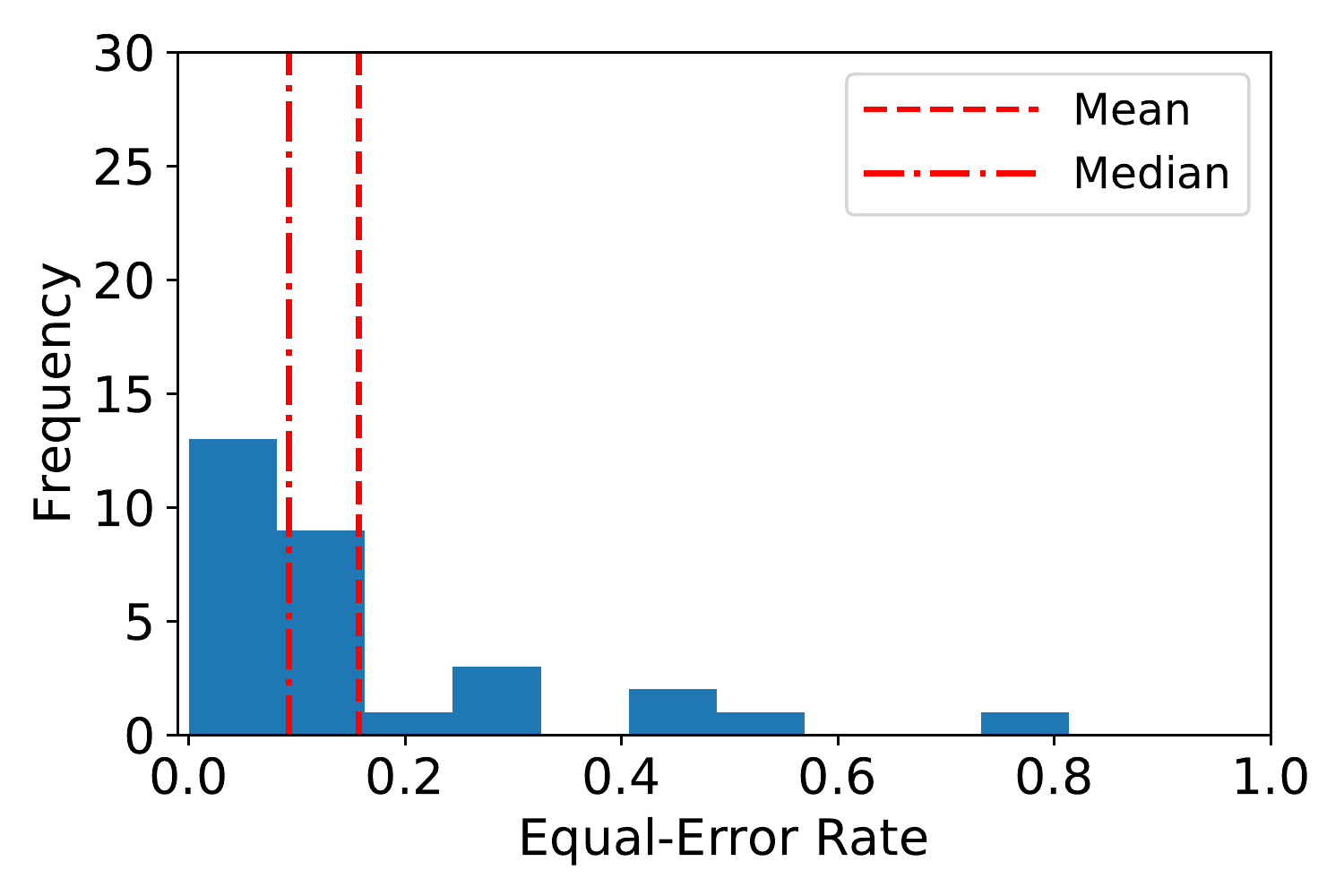}%
        \label{subfig_inf_mix_ots_imp_score_dist}}
    \end{center}
    \caption{Equal-Error Rate Distributions}
    \label{fig_dataset_dists}
\end{figure}

The experimental results are shown in Table \ref{tab_results}. Accompanying plots of the score distributions with the mean and median overlaid are shown in Figure \ref{fig_dataset_dists}.

The Bayesian Gaussian model performed the best in the blind impostor case, with a mean EER of 4.54\%. All of the models performed quite well in this scenario, however, as each model classified the majority of the samples correctly. This can be seen from the left-hand column of Figure \ref{fig_dataset_dists}. It was expected that the models would perform well in the blind impostor scenario, as it is unlikely that an uninformed attacker would be able to accurately replicate the the genuine user's swipe gesture.

The error rates for the OTS impostor experiments are somewhat higher than those of the blind impostors, this is again expected due to the attacker now having information about the genuine user's gestures. The models still remain quite discriminative, however, with the infinite mixture model having the best overall mean EER of 15.70\%. The Bayesian Gaussian model has arguably the best distribution of scores in this scenario, with a median EER of 5.96\%. A larger-scale experiment would be required to definitively establish which of these models is the overall best-performing.

\begin{figure}[!t]
    \begin{center}
        \subfloat[Good Behaviour Separability][Good Behaviour Separability]{\includegraphics[width=0.5\linewidth,trim={0 1.0in 0 1.2in},clip]{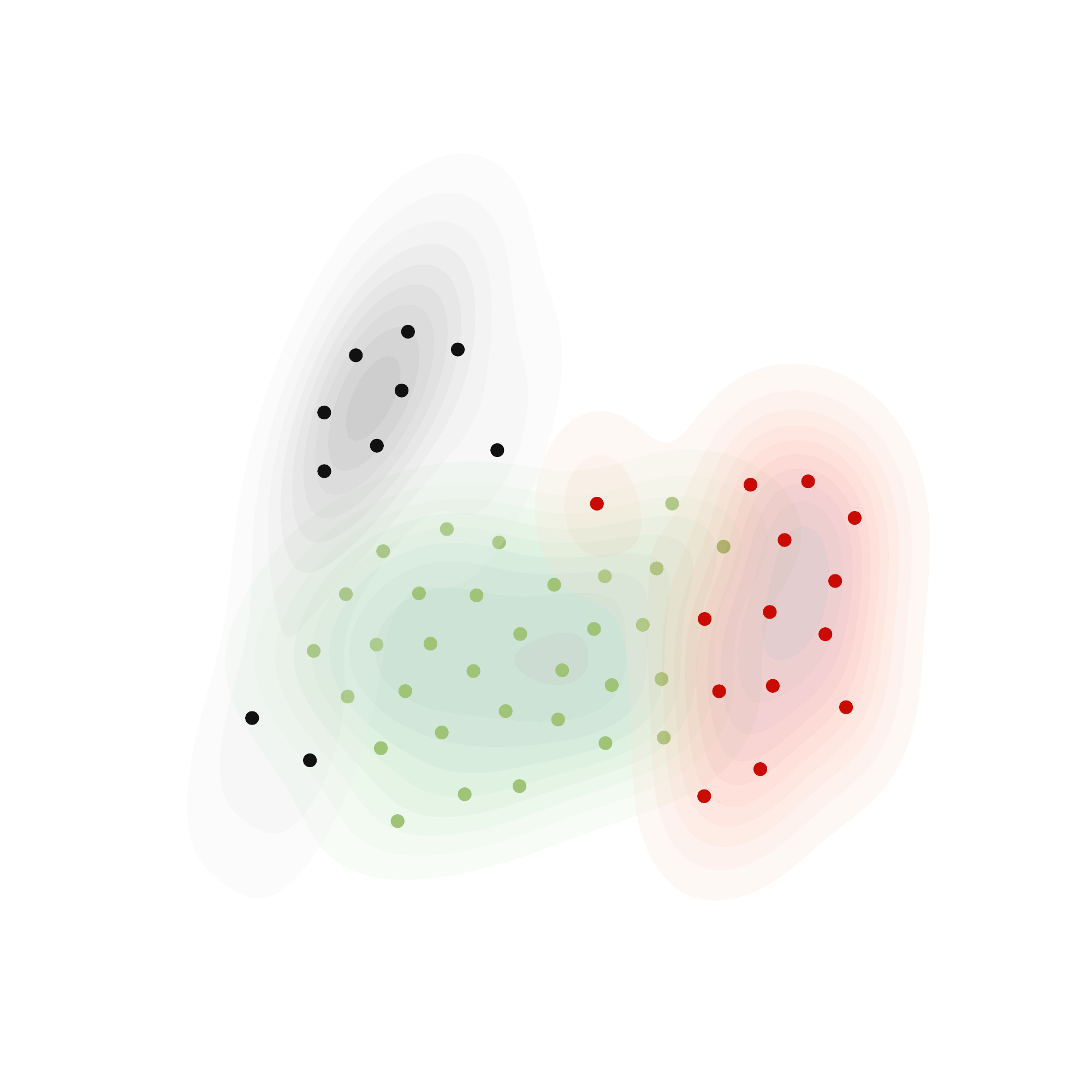}%
        \label{subfig_tsne_good_separability}}
        \hfil
        \subfloat[Bad Behaviour Separability][Bad Behaviour Separability]{\includegraphics[width=0.5\linewidth,trim={0 2.0in 0 2.0in},clip]{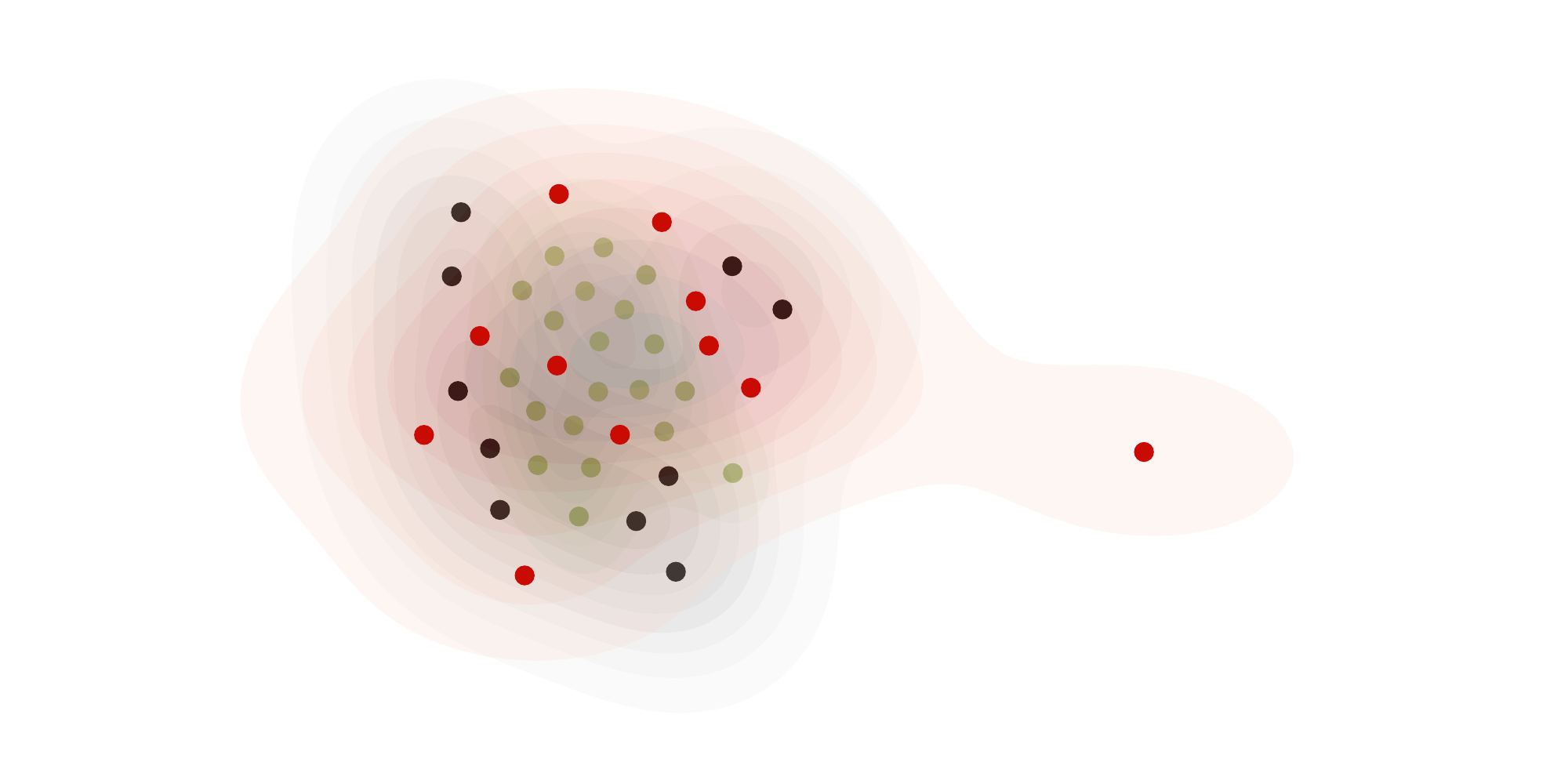}%
        \label{subfig_tsne_bad_separability}}
    \end{center}
    \caption{Illustration of victim (green), blind (black) and OTS (red) swipes using a t-SNE projection into a 2-dimensional space.}
    \label{fig_tsne_separability}
\end{figure}

These results show that it is generally difficult to precisely replicate another user's behaviour even with knowledge of the required gestures. There is an approximate 10\% increase in EERs between the blind and OTS impostor sessions. Considering the large feature space it is unsurprising that an attacker has difficulty replicating the correct behaviour across all the features. Figure \ref{fig_tsne_separability} shows two sample profiles with the features projected into a 2-dimensional space, one with easily separable behaviour and the other without. The example with good separability had low EERs across all the experiments; this is relatively easy to see as the impostor samples are well-separated from the genuine samples. The other example was the worst performing profile in the dataset, equally the reason for this is readily apparent as all of the distributions are overlapping. Note that some information is lost when the features are projected into 2-dimensional space, these plots are merely for illustration of the problem.

\subsection{Multiple Behaviours}
\begin{figure}[t]
    \centering
    \includegraphics[width=0.6\linewidth]{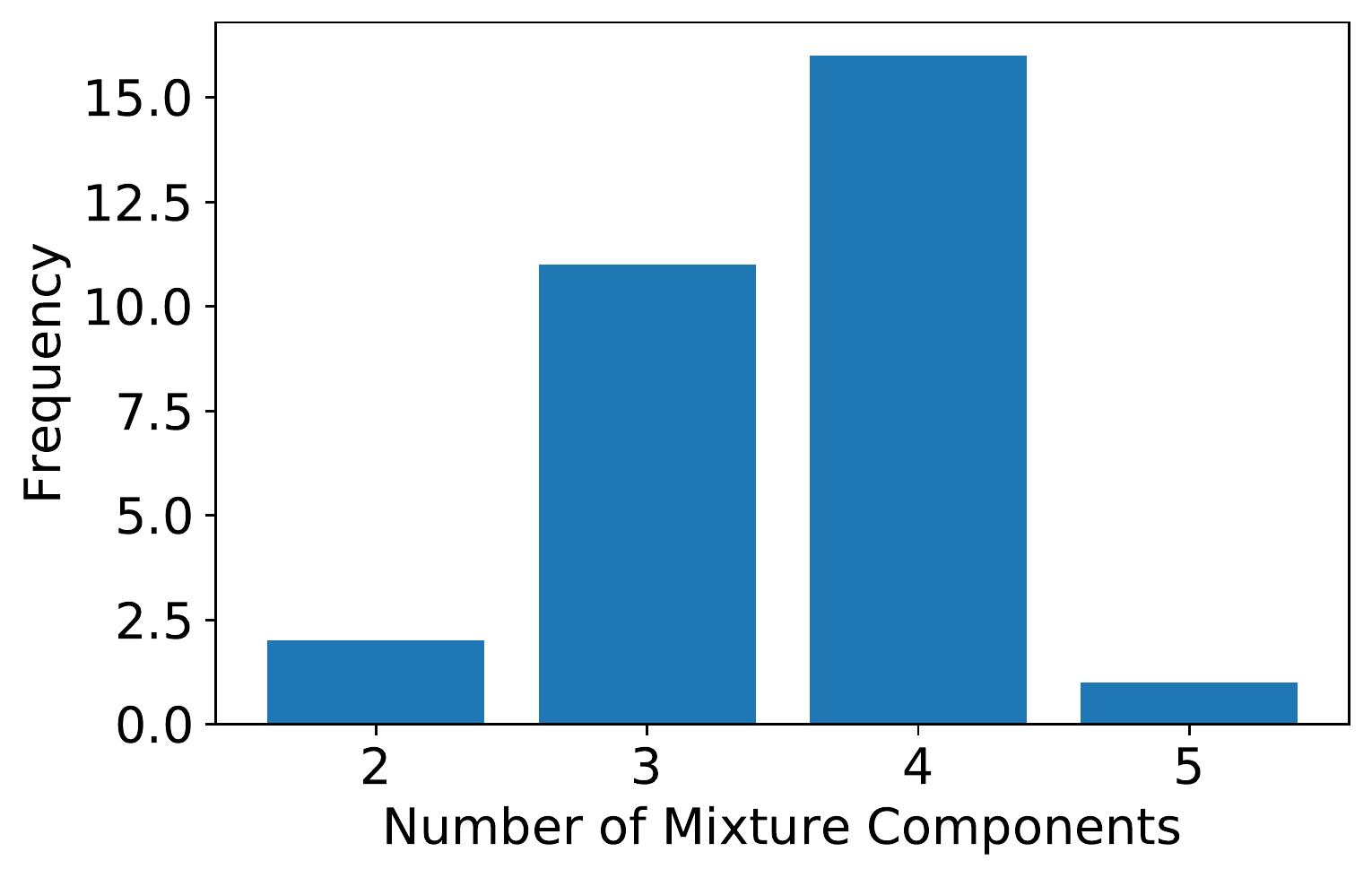}
    \caption{Infinite Mixture Component Usage Distribution}
    \label{fig_inf_mixture_comps}
\end{figure}

The infinite mixture model has the inherent capability to learn multiple distributions from the data. Though the training sample sizes used in the experiments are somewhat small for explicit development of multiple behaviours, the mixture model learned multiple distributions for the profiles regardless. Figure \ref{fig_inf_mixture_comps} shows the distribution of the number of mixture components learned per profile. The somewhat large number of distributions is possibly due to the spherical priors used for the component covariances which have encouraged the model to learn feature correlations via the components. This does not appear to have unduly affected the model performance, though the phenomenon bears further research.

\subsection{Low-Sample Learning}

\begin{figure}
    \captionsetup[subfigure]{justification=centering}
    \captionsetup[subfloat]{farskip=2pt}
    \begin{center}
        \subfloat[Mean EER, Blind Impostor][Mean EER, Blind Impostor]{\includegraphics[width=0.5\linewidth]{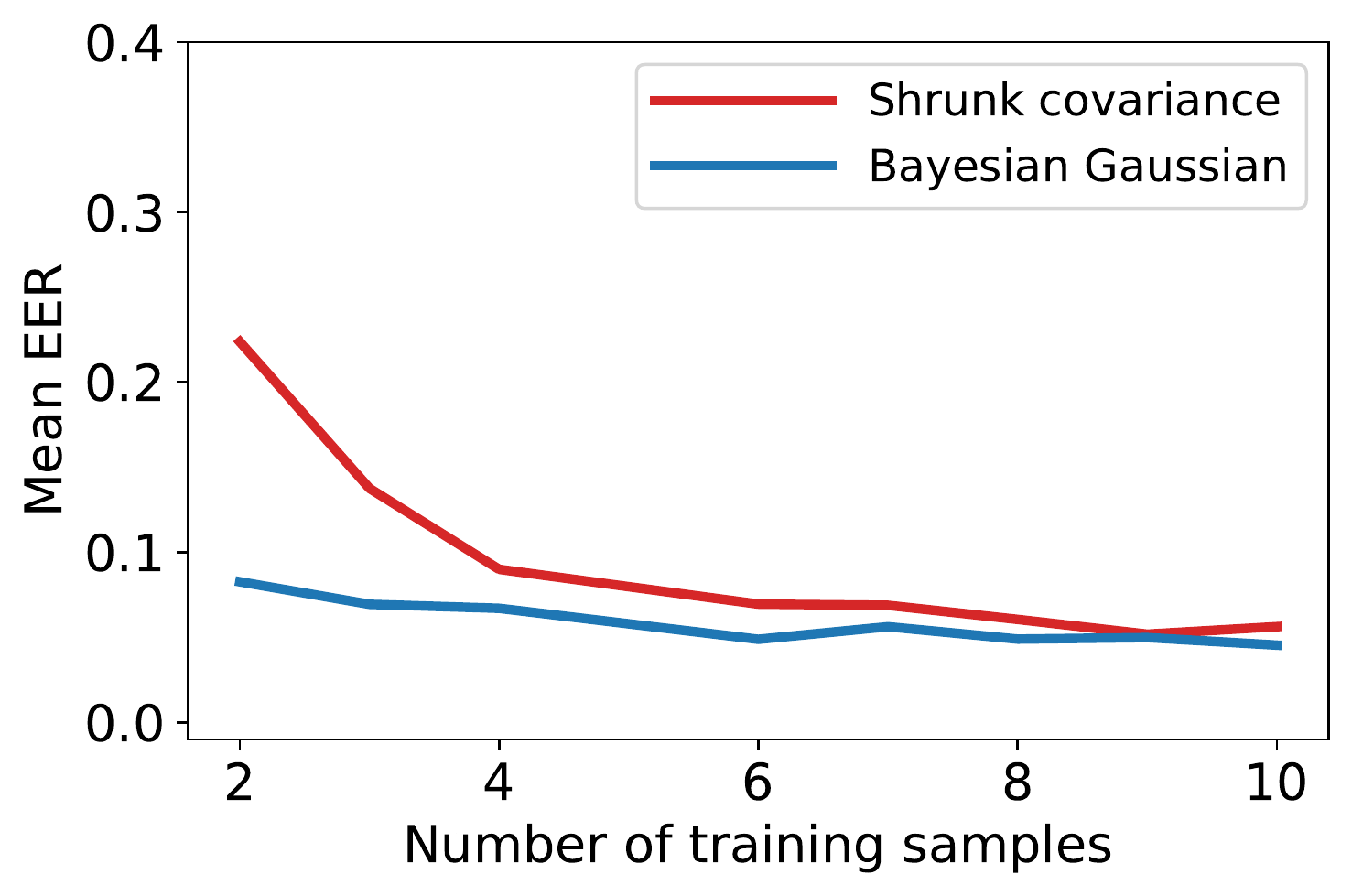}%
        \label{subfig_mean_blind_learning_curve}}
        \hfil
        \subfloat[Mean EER, OTS Impostor][Mean EER, OTS Impostor]{\includegraphics[width=0.5\linewidth]{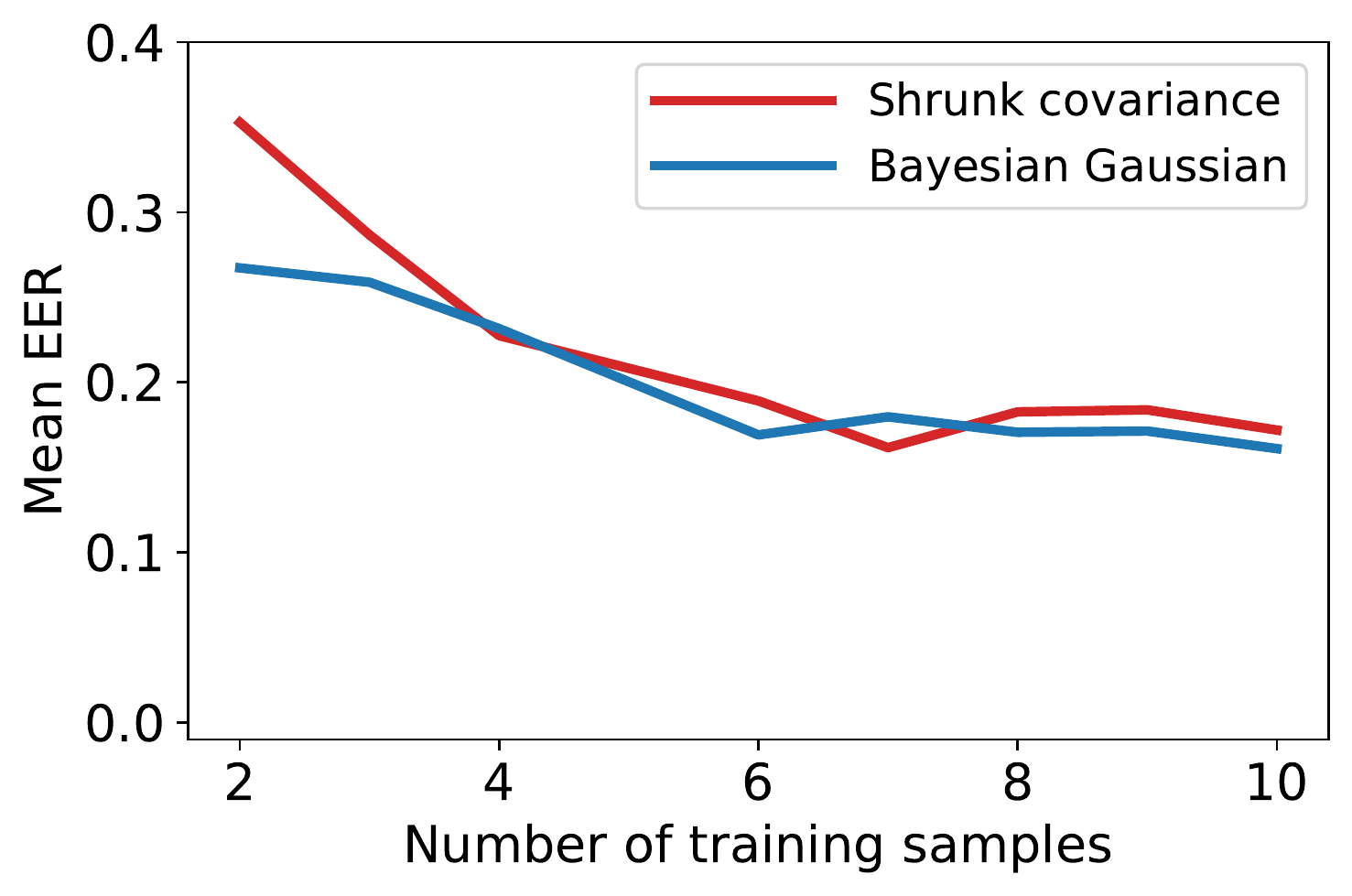}%
        \label{subfig_mean_ots_learning_curve}}\\
        \subfloat[Median EER, Blind Impostor][Median EER, Blind Impostor]{\includegraphics[width=0.5\linewidth]{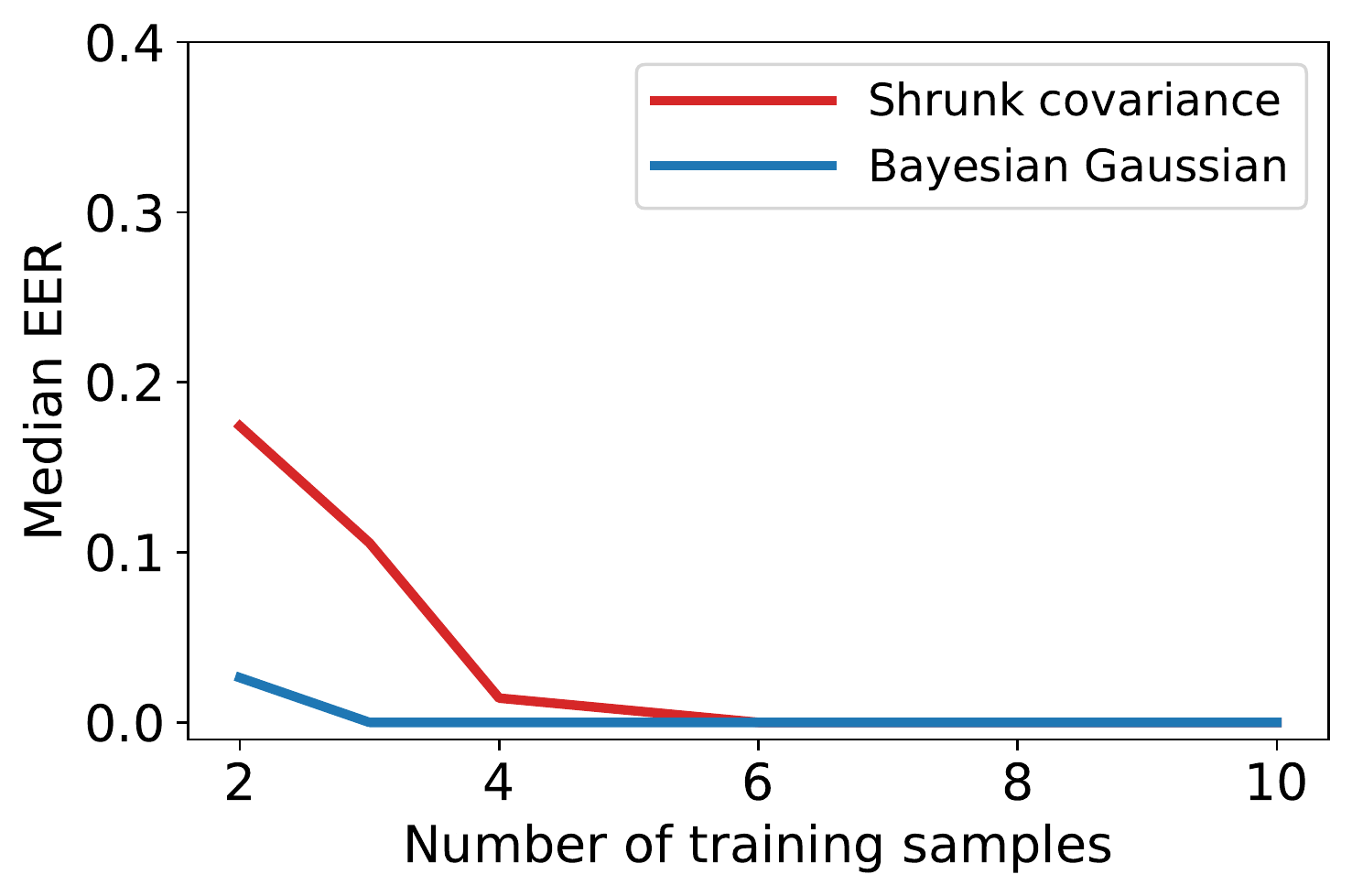}%
        \label{subfig_median_blind_learning_curve}}
        \hfil
        \subfloat[Median EER, OTS Impostor][Median EER, OTS Impostor]{\includegraphics[width=0.5\linewidth]{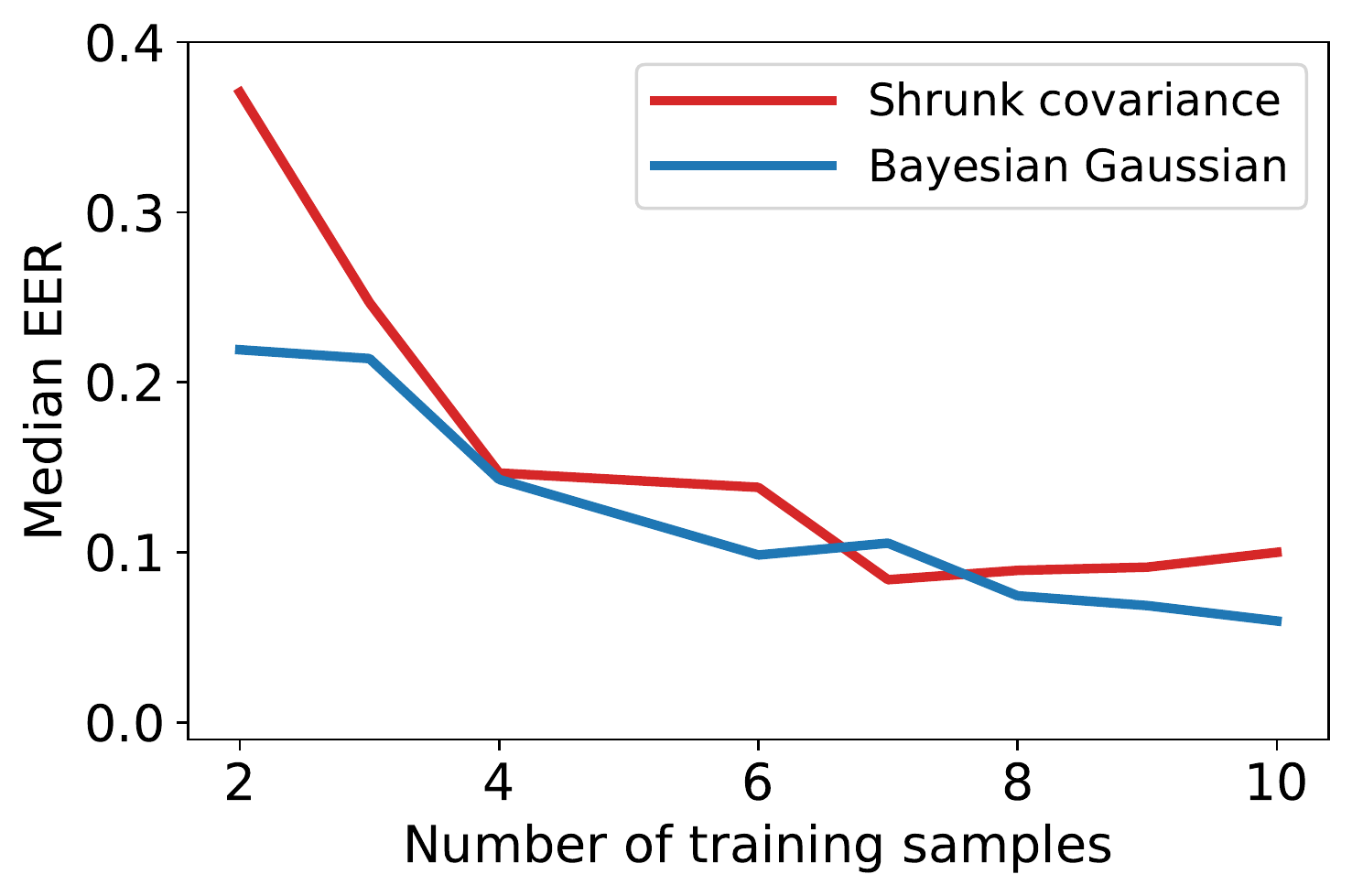}%
        \label{subfig_median_ots_learning_curve}}
    \end{center}
    \caption{Learning Curves; Shrunk Covariance and Bayesian Gaussian Models}
    \label{fig_learning_curves}
\end{figure}

In Figure \ref{fig_learning_curves}, learning curves for the shrunk covariance and Bayesian Gaussian models are shown, from 2 to 10 training samples. These plots show the key advantage of using a Bayesian model when few training samples are available. The Bayesian model shows significantly lower EERs than the shrunk covariance model when only 2-5 training samples are available. The effect is more pronounced in the blind impostor scenario as the data is generally easier to learn, as established above. To a somewhat lesser degree, the benefit is still observed in the OTS impostor scenario.

From these results it is apparent that a Bayesian approach is highly effective for the swipe dynamics problem as it enables accurate predictions in high-dimensional data with only a few training samples.

\section{Conclusion and Future Work}
This paper has introduced and motivated an unsupervised Bayesian approach to non-continuous swipe dynamics, a type of behavioural biometric. Specifically, some of the key problems facing this type of authentication, particularly around accurate data collection and quality, have been discussed. The lack of large scale datasets in this domain has driven recent research efforts to the use of unsupervised learning models.

Three different types of probabilistic models were introduced: a Gaussian shrunk covariance, a Bayesian multivariate Gaussian, and an infinite Gaussian mixture. Two different attacker scenarios are presented, those of a blind and and over-the-shoulder impostor. The three models are compared across the different attack scenarios, with the Bayesian Gaussian model showing the best performance with a mean EER of 4.54\% for the blind impostor scenario and the infinite mixture model for the OTS impostor with a mean EER of 15.70\%. Recursive training results are shown for the Bayesian Gaussian and shrunk covariance models, where the Bayesian model converges to the final error rate with approximately 30\% of the training samples required by the shrunk covariance in some cases. This indicates the strength of the Bayesian approach in general.

There are two key areas of research that bear further investigation. First, the use of hierarchical priors for the Bayesian models presented in this paper. The priors used here are constructed relatively naively and it is likely that the model performance could be improved, for example, through incorporating information from a population of similar users or mobile devices. Secondly, modelling multiple behaviours has only been investigated briefly in this paper and more in-depth research is warranted. Specifically, curating a dataset with explicit labelled single- and multiple-behaviour profiles and comparing the modelling approaches presented here with others is a logical next step.

{\small
\bibliographystyle{ieee}
\bibliography{bibliography}
}

\end{document}